# REVIEW OF $\alpha_s$ MEASUREMENTS[*]

P.N. Burrows[**]

*Stanford Linear Accelerator Center*
*Stanford University, Stanford, CA94309, USA*

Email: burrows@slac.stanford.edu

## ABSTRACT

Determinations of $\alpha_s$ are reviewed. Current results are limited to a precision of around 3-20%, largely by theoretical uncertainties. All measurements are consistent with a 'world average' value of $0.118 \pm 0.005$ and there is no evidence of any discrepancy between 'low-$Q^2$' and 'high-$Q^2$' results.

Talk presented at the
3rd International Symposium on Radiative Corrections
August 1-5 1996, Cracow, Poland

[*] Work supported by Department of Energy contracts DE–AC02–76ER03069 (MIT) and DE–AC03–76SF00515 (SLAC).
[**] Permanent address: Laboratory for Nuclear Science, Massachusetts Institute of Technology, Cambridge, MA 02139, USA.

# 1 Introduction

One of the most important areas of high energy physics comprises tests of the theory of strong interactions, Quantum Chromodynamics (QCD) [1]. Since the theory contains in principle only one free parameter, the strong interaction scale $\Lambda$, tests of QCD can be quantified in terms of comparison of measurements of $\Lambda$ in different processes and at different hard scales $Q$. In practice most QCD calculations of observables are performed using finite-order perturbation theory, and calculations beyond leading order depend on the *renormalisation scheme* employed, implying a scheme-dependent $\Lambda$. It is conventional to work in the modified minimal subtraction scheme ($\overline{MS}$ scheme) [2], and to use the strong interaction scale $\Lambda_{\overline{MS}}$ for five active quark flavours. If one knows $\Lambda_{\overline{MS}}$ one may calculate the strong coupling $\alpha_s(Q^2)$ from the solution of the QCD renormalisation group equation [3]. Because of the large data samples taken in e$^+$e$^-$ annihilation at the $Z^0$ resonance, it has become conventional to use as a yardstick $\alpha_s(M_Z^2)$, where $M_Z$ is the mass of the $Z^0$ boson; $M_Z \approx 91.2$ GeV [4]. Tests of QCD can therefore be quantified in terms of the consistency of the values of $\alpha_s(M_Z^2)$ measured in different experiments; such measurements have been performed in e$^+$e$^-$ annihilation, hadron-hadron collisions, and deep-inelastic lepton-hadron scattering, covering a range of $Q^2$ from roughly 1 to $10^5$ GeV$^2$. In addition to testing QCD, the precise measurement of $\alpha_s(M_Z^2)$ allows constraints on possible extensions to the Standard Model (SM) of elementary particles; see *eg.* [5].

Over the past decade many measurements of $\alpha_s$ have been presented. These are reviewed here, with emphasis on the more recent results. The systematic uncertainties associated with these measurements, both experimental and theoretical in nature, are discussed, and the degree of consistency between $\alpha_s(M_Z^2)$ values is examined. The task of preparing this article has been aided by a number of previous reviews [3, 6].

# 2 Theoretical Considerations

An inclusive observable $X$ may be written schematically:

$$X = X^{EW} (1 + \delta^{QCD}) \qquad (1)$$

where $X^{EW}$ represents the electroweak contribution. Since, with observables of this type, $\alpha_s$ enters via the small QCD radiative correction, $\delta^{QCD}$, a precise measurement of $\alpha_s$ generally requires a large data sample. Observables can also be defined that are directly proportional to $\delta^{QCD}$ and hence potentially more sensitive to $\alpha_s$. In either case $\delta^{QCD}$ can be separated into perturbative and non-perturbative contributions:

$$\delta^{QCD} = \delta^{pert} + \delta^{non-pert}. \tag{2}$$

The perturbative contribution can in principle be calculated as a power series in $\alpha_s$, though in practice the large number of Feynman diagrams involved renders calculations beyond the first few orders intractable. An observable must be calculated to at least next-to-leading order in the $\overline{MS}$ scheme so as to define $\Lambda_{\overline{MS}}$; the solution of the renormalisation group equation [3] to the same order can then be used to translate consistently to the yardstick $\alpha_s(M_Z^2)$. The non-perturbative contribution, often called a 'hadronisation correction' in e$^+$e$^-$ annihilation or a 'higher twist effect' in lepton-hadron scattering, is expected [7] to have the form of a series of inverse powers of the physical scale. Hence

$$\delta^{QCD} = \Sigma_i a_i \left(\frac{\alpha_s}{\pi}\right)^i + \Sigma_j \frac{b_j}{Q^j}. \tag{3}$$

In general, due to the presence of 'renormalon ambiguities' in perturbation theory [7], the perturbative contribution cannot be calculated independently of consideration of the non-perturbative power-law contribution. For several inclusive observables the perturbative series has been calculated to next-to-next-to-leading order; for jet-like observables it has been calculated to next-to-leading order. The power-law corrections cannot in general be calculated, though there has been recent progress in this direction [7]. Therefore, in any comparison of a QCD prediction with data, the uncertainties relating to both the uncalculated higher-order perturbative, as well as non-perturbative, contributions should be estimated, and a *theoretical uncertainty* on the extracted value of $\alpha_s(M_Z^2)$ assigned accordingly.

From an operational point-of-view, truncation of the perturbative series at finite order causes a residual dependence on the scheme-dependent *renormalisation scale* $\mu$. This parameter is formally unphysical and should not enter at all into an exact infinite-order calculation, and its value is arbitrary. There is some, but by no means universal, consensus that the effect of missing higher-order terms can hence be estimated from the dependence of $\alpha_s(M_Z^2)$ on the value of $\mu$ assumed in fits of the calculations to the data, and a *renormalisation scale uncertainty* is sometimes quoted. This procedure, well-motivated in that the $\mu$-dependence caused by the truncation of the perturbation series would be cancelled by addition of the higher-order terms, is, however, arbitrary, and is not equivalent to knowledge of the size of the *a priori* unknown terms. In cases where scale uncertainties are considered this arbitrariness is manifested in the wide variation among the ranges and central values of $\mu$ chosen by different experimental groups, see *eg.* [8]; in other cases this source of uncertainty is not included in the errors. Different $\alpha_s(M_Z^2)$ results with similar experimental precision can hence be quoted with different total errors depending on the procedure adopted for assigning the theoretical uncertainties. The interpretation of the central values and errors on $\alpha_s(M_Z^2)$ measurements is hence not always straightforward. In this review it is attempted to summarise the sources of error and assumptions made in the various determinations.

## 3  e$^+$e$^-$ Annihilation

Since hadronic activity is restricted to the final state, the simplest environment for measurement of $\alpha_s$ is provided by e$^+$e$^-$ annihilation. Experimental signatures of hadronic events are largely free of backgrounds, and the smaller number of Feynman diagrams contributing at a given order in perturbation theory renders QCD calculations more tractable than in lepton-hadron or hadron-hadron collisions.

### 3.1  $R$ and $Z^0$ Lineshape

For the inclusive ratio $R = \sigma(\text{e}^+\text{e}^- \to \text{hadrons})/\sigma(\text{e}^+\text{e}^- \to \mu^+\mu^-)$, the SM electroweak contributions are well understood theoretically and the perturbative QCD series has been calculated up to O($\alpha_s^3$) [9] for massless quarks and up to O($\alpha_s^2$) including quark mass effects [10]; the large size of the O($\alpha_s^3$) term is potentially a cause for concern about the degree of convergence of the series. The SM prediction for $R$ with the top quark mass fixed to 179 GeV, $M_Z$ fixed to 91.185 GeV, and the SM Higgs mass $M_H$ fixed to 300 GeV was fitted to published measurements below the $Z^0$ resonance, in the c.m. energy range $5 \leq Q \leq 65$ GeV, and yielded [11]:

$$\alpha_s(M_Z^2) \;\; = \;\; 0.128^{+0.012}_{-0.013}(\text{exp.}) \pm 0.002(M_H). \tag{4}$$

The experimental error is large due to the small data samples and/or limited precision of the luminosity measurements. The second error has been assigned for variation of $M_H$ between 60 and 1000 GeV.

The CLEO Collaboration is currently attempting to measure $R$ at $Q = 10.5$ GeV. They expect to achieve a precision of order $\Delta R \sim \pm 2.9\%$ [12], which will translate into a precision of about $\pm 0.04$ on $\alpha_s(M_Z^2)$, providing a useful, but not very precise, low-energy measurement.

Closely-related observables at the $Z^0$ resonance are the $Z^0$ total width $\Gamma_Z$, the pole cross section $\sigma_h^0$, and the ratio of leptonic to hadronic $Z^0$ decay branching ratios $R_l$, which have been measured using the sample of approximately 16M hadronic $Z^0$ decays measured at LEP. In these cases the non-perturbative contributions are expected to be $O(1/M_Z)$ and are usually ignored. A concern is that recent measurements of observables that probe the electroweak couplings of the $Z^0$ to b and c quarks deviate slightly from SM expectations [13]. Since these couplings must be known in order to extract $\alpha_s(M_Z^2)$, this effect, whatever its origin, is a potential source of bias [3]. Further analysis is in progress from the SLC and LEP experiments and the situation is not yet resolved.

Proceeding nonetheless, the procedure adopted [13] is to perform a global fit to a panoply of electroweak data that includes the W and top quark masses as well as the $Z^0$ observables relating to the lineshape, left-right production asymmetry, decay fermion forward-backward asymmetries, branching ratios to heavy quarks, and $\tau$ polarisation, by allowing $M_H$ and $\alpha_s(M_Z^2)$ to vary. Data presented at the 1996 summer conferences yield the positively-correlated results [13] $M_H = 149^{+190}_{-82}$ GeV and

$$\alpha_s(M_Z^2) = 0.1202 \pm 0.0033 (\text{exp.}). \qquad (5)$$

The $\alpha_s(M_Z^2)$ value is lower than the corresponding results presented at the 1995 conferences [14], $\alpha_s(M_Z^2) = 0.123 \pm 0.005$, and at the 1994 conferences, $\alpha_s(M_Z^2)$ $0.125 \pm 0.005$ [15], whose large central values were partly responsible for a supposed discrepancy between 'low-$Q^2$' and 'high-$Q^2$' $\alpha_s(M_Z^2)$ measurements [16]. The change between 1995 and 1996 is due to a combination of shifts in the values of the $Z^0$ lineshape parameters, redetermined in light of the recalibration of the LEP beam energy due to the 'TGV effect' [13], and a change in the central value of $M_H$ at which $\alpha_s(M_Z^2)$ is quoted, from 300 GeV (1995) to the fitted value 149 GeV (1996). A detailed study of theoretical uncertainties implies [17] that they contribute at a level substantially below $\pm 0.001$. Since data-taking at the $Z^0$ resonance has now been completed at the LEP collider the precision of this result is not expected to improve further.

## 3.2 τ Decays

An inclusive quantity similar to $R$ is the ratio $R_\tau$ of hadronic to leptonic decay branching ratios, $B_h$ and $B_l$ respectively, of the $\tau$ lepton:

$$R_\tau \equiv \frac{B_h}{B_l} = \frac{1 - B_e - B_\mu}{B_e} \qquad (6)$$

where $B_e$ and $B_\mu$ can either be measured directly, or deduced from a measurement of the $\tau$ lifetime $\tau_\tau$. In addition, a family of observables known as 'spectral moments' of the invariant mass-squared $s$ of the hadronic system has been proposed [18]:

$$R_\tau^{kl} \equiv \frac{1}{B_e} \int_0^{M_\tau^2} ds \left(1 - \frac{s}{M_\tau^2}\right)^k \left(\frac{s}{M_\tau^2}\right)^l \frac{dB_h}{ds} \qquad (7)$$

where $M_\tau$ is the $\tau$ mass. In this case the integrand can be measured independently of $B_e$. It is easily seen that $R_\tau = R_\tau^{00}$.

$R_\tau$ and $R_\tau^{kl}$ have been calculated perturbatively up to $O(\alpha_s^3)$. However, because $M_\tau \sim 1$ GeV one expects $\alpha_s(M_\tau) \sim 0.3$ and it is not *a priori* obvious that the perturbative calculation can be expected to be reliable, or that the non-perturbative contributions will be small. In recent years a large theoretical effort has been devoted to this subject; see *eg.* [18, 19, 20].

The ALEPH Collaboration derived $R_\tau$ from its own measurements of $B_e$, $B_\mu$, and $\tau_\tau$, and also measured the (10), (11), (12), and (13) spectral moments. A combined fit yielded [21] $\alpha_s(M_\tau) = 0.330 \pm 0.046$ ($\alpha_s(M_Z^2) = 0.118 \pm 0.005$) and $\delta^{non-pert} = 0.003 \pm 0.005$. A later conference report [22] quoted $\alpha_s(M_Z^2) = 0.124 \pm 0.0022 \pm 0.001$, where the first error receives equal contributions from experiment and theory, and the second derives from uncertainties in evolving $\alpha_s$ across the c and b thresholds; this result will be included in the average presented here. The OPAL Collaboration measured $R_\tau$ from $B_e$, $B_\mu$, and $\tau_\tau$, and derived [23] $\alpha_s(M_\tau) = 0.375^{+0.019}_{-0.018}$ (exp.) $^{+0.025}_{-0.017}$ (pert.) $\pm 0.006$ (non-pert.) ($\alpha_s(M_Z^2) = 0.1229^{+0.0016}_{-0.0017}$ (exp.) $^{+0.0025}_{-0.0021}$ (theor.)). The CLEO Collaboration measured the same four spectral moments as ALEPH and also derived $R_\tau$ using 1994 Particle Data Group values for $B_e$, $B_\mu$ and $\tau_\tau$. A combined fit yielded [24] $\alpha_s(M_\tau) = 0.306 \pm 0.017$ (exp.) $\pm 0.017$ (theor.) ($\alpha_s(M_Z^2) = 0.114 \pm 0.003$). This central value is slightly lower than the ALEPH and OPAL values. If more recent world average values of $B_e$ and $B_\mu$ are used CLEO obtains a central $\alpha_s(M_\tau)$ value of 0.339 [24]. Averaging the second CLEO result and the ALEPH and OPAL results by weighting with the experimental errors, assuming they are uncorrelated, yields $\alpha_s(M_Z^2) = 0.122 \pm 0.001$ (exp.) $\pm 0.002$ (theor.). This is nominally a very precise measurement.

Recently, however, theoretical studies have been performed which suggest that additional uncertainties need to be added to the above result before it can be compared with other measurements of $\alpha_s(M_Z^2)$. A comparison of the exact $O(\alpha_s^3)$ perturbative calculation of the spectral moments with a resummed procedure and with a partial resummation based on 'renormalon chains' concluded [20] that the theoretical uncertainty on $\alpha_s(M_\tau)$ is $\pm 0.05$. A study of ultraviolet renormalon ambiguities in the behaviour of the perturbative series yielded [25] an estimate of the theoretical uncertainty on $\alpha_s(M_\tau)$ of at least $\pm 0.06$. Until there is a resolution of these theoretical considerations it seems appropriate to conclude that the current value of $\alpha_s(M_Z^2)$ from $\tau$ decays is:

$$\alpha_s(M_Z^2) = 0.122 \pm 0.001 (\text{exp.}) \pm 0.006 (\text{theor.}) \qquad (8)$$

## 3.3 Hadronic Event Shape Observables

In e$^+$e$^-$ annihilation one can define infra-red- and collinear-safe measures of the topology of hadronic final states. Perhaps the most familiar such quantity is the rate of production of 3-jet events, defined using an iterative particle clustering algorithm, see *eg.* [26], but there are other measures relating to longitudinal momentum flow in events, jet masses, and energy-energy correlations between particles; for a discussion see *eg.* [27]. The observables are constructed to be directly proportional to $\alpha_s$ at leading order, and so are potentially sensitive measures of the strong coupling. To date the leading and next-to-leading order terms have been calculated for a large number of observables [28].

The technology of this approach has been developed over the past 15 years of analysis at the PETRA, PEP, TRISTAN, SLC and LEP colliders, so that the method is considered to be well understood both experimentally and theoretically. It is necessary to correct the measured distributions for any bias effects originating from the detector acceptance, resolution, and inefficiency, as well as for the effects of initial-state radiation and hadronisation, to yield 'parton-level' distributions which can be compared directly with the QCD calculations. In the absence of non-perturbative calculations of the hadronisation process, Monte Carlo models of jet fragmentation [29, 30] are commonly used for this purpose.

The $Z^0$ experiments have measured $\alpha_s(M_Z^2)$ in this fashion; see [8] and references therein. For example, the recent SLD study employed 15 observables and found considerable scatter among the 15 $\alpha_s(M_Z^2)$ values, the r.m.s. deviation being 0.0076. This scatter is much larger than the hadronisation uncertainties and can be interpreted as arising from the effects of the uncalculated higher-order perturbative contributions. In fact, within the renormalisation scale uncertainties the $\alpha_s(M_Z^2)$ values from all the observables are consistent. The experimental collaborations have chosen different sets of obervables and different ranges over which to vary $\mu$; in combination with their different averaging methods this has led to some variation among the quoted central values of $\alpha_s(M_Z^2)$ and scale uncertainties [8].

For six observables improved calculations can be formulated that incorporate the resummation [31] of leading and next-to-leading logarithmic terms matched to the O($\alpha_s^2$) results. The matched calculations are expected *a priori* both to describe the data in a larger region of phase space than the fixed-order results, and to yield a reduced dependence of $\alpha_s$ on the renormalization scale; they have been applied by the $Z^0$ experiments to determine $\alpha_s(M_Z^2)$ [27, 32].

Hinchliffe has reviewed the various measurements from experiments performed in the c.m. energy range $10 \leq Q \leq 91$ GeV, utlilising both O($\alpha_s^2$) and resummed calculations, and quotes an average value of $\alpha_s(M_Z^2) = 0.122 \pm 0.007$ [3], where the large error is dominated by the scale uncertainty, which far exceeds the experimental error of about $\pm 0.002$. Schmelling has also compiled the measurements, including the recent results from the LEP run at $Q \sim 133$ GeV [33], and quotes a global average [34] $\alpha_s(M_Z^2) = 0.121 \pm 0.005$, in agreement with [3], but assuming a more aggressive scale uncertainty. The value

$$\alpha_s(M_Z^2) \quad = \quad 0.121 \pm 0.002(\text{exp.}) \pm 0.005(\text{theor.}) \qquad (9)$$

will be used here.

The best way to reduce the theoretical uncertainty would be to calculate the observables to higher order in perturbation theory, a difficult task that has not yet been performed. In the absence of O($\alpha_s^3$) QCD calculations it has been suggested [35] that the O($\alpha_s^2$) calculation for each observable can be 'optimised' by choosing a specific value of the renormalisation scale. This approach has recently been applied [8] to 15 event shape observables using the 'PMS', 'FAC', and 'BLM' optimised scales. However, for any of these scale choices the scatter among the $\alpha_s(M_Z^2)$ values is comparable with that from the choice $\mu = Q$, implying that higher-order effects contribute roughly equally in all of these procedures. Recently Padé Approximants have been applied to estimate the O($\alpha_s^3$) contributions to the QCD series for the same 15 event shape observables [36]. Remarkably the scatter among the $\alpha_s(M_Z^2)$ values is noticeably smaller than in the O($\alpha_s^2$) cases, the r.m.s. deviation being ±0.0035. This convergence implies that the Padé method provides at least a partial approximation of higher-order perturbative QCD contributions, but explicit calculation of the O($\alpha_s^3$) terms will be necessary to confirm this.

Finally, the L3 Collaboration has utilised events with a hard radiated final-state photon, which reduces the effective c.m. energy available to the hadronic system, to examine the $Q^2$-evolution of four event shape observables in the range $30 \leq Q \leq 86$ GeV. By comparing with resummed + O($\alpha_s^2$) calculations they derived [37] preliminary values of $\alpha_s$ in each energy bin that, when evolved to $M_Z$, are consistent with $\alpha_s(M_Z^2) = 0.120$.

## 3.4 Scaling Violations in Fragmentation Functions

Though distributions of final-state hadrons are not, in general, calculable in perturbative QCD, the $Q^2$-evolution of the scaled energy ($x_p = 2E/Q$) distributions of hadrons, or 'fragmentation functions', can be calculated and used to determine $\alpha_s$. In addition to the usual renormalisation scale $\mu$, a *factorisation scale* $\mu_F$ must be defined that delineates the boundary between the calculable perturbative, and incalculable non-perturbative, domains. Additional complications arise from the changing composition of the underlying event flavour with $Q$ due to the different $Q$-dependence of the $\gamma$ and $Z^0$ exchange processes. Since B and C hadrons typically carry a large fraction of the beam momentum, and contribute a large multiplicity from their decays, it is necessary to consider the scaling violations separately in b, c, and light quark events, as well as in gluon jet fragmentation.

In an early analysis [38] the DELPHI Collaboration parametrised the fragmentation functions using the O($\alpha_s^2$) matrix elements and the string fragmentation model implemented in JETSET [29]. They fitted data in the range $14 \leq Q \leq 91$ GeV to determine $\alpha_s(M_Z^2) = 0.118 \pm 0.005$, where the error is dominated by varying $\mu$ in the range $0.1 \leq \mu/Q \leq 1$. The ALEPH Collaboration used its $Z^0$ data to constrain flavour-dependent effects by tagging event samples enriched in light, c, and b quarks, as well as a sample of gluon jets [39]. The fragmentation functions for the different flavours and the gluon were parametrised at a reference energy, evolved with $Q$ according to the perturbative DGLAP formalism calculated at next-to-leading order [40], in conjunction with a parametrisation proportional to $1/Q$ to represent non-perturbative effects [41], and fitted to data in the range $22 \leq Q \leq 91$ GeV. They derived $\alpha_s(M_Z^2) = 0.126 \pm 0.007$ (exp.) $\pm 0.006$ (theor.), where the theoretical uncertainty is dominated by variation of the factorisation scale $\mu_F$ in the range $-1 \leq \ln\mu_F^2/Q^2 \leq 1$; variation of the renormalisation scale in the same range contributed only $\pm 0.002$. DELPHI has recently reported a similar analysis [42] yielding $\alpha_s(M_Z^2) = 0.121^{+0.006}_{-0.007}$ (exp.) $\pm 0.010$ (theor.). Curiously, although a similar range as ALEPH, $0.3 \leq \mu/Q \leq 3$, was used to examine variation of the renormalisation and factorisation scales, here the renormalisation scale dominates the theoretical uncertainty, with a contribution of $\pm 0.009$, in contrast to $\pm 0.002$ from factorisation. Combining the ALEPH and later DELPHI results, assuming uncorrelated experimental errors, yields:

$$\alpha_s(M_Z^2) = 0.124 \pm 0.005(\text{exp.}) \pm 0.010(\text{theor.}) \qquad (10)$$

where, until the apparent discrepancy between the ALEPH and DELPHI analyses of theoretical uncertainties is resolved, the larger DELPHI value is taken.

## 4 Lepton-hadron Scattering

The study of scaling violations in structure functions has been of historical significance in establishing QCD as the theory of strong interactions. Measurements have been performed using electron, muon and neutrino beams directed onto a variety of both polarised and unpolarised nuclear targets. An enormous range of the $(x, Q^2)$ kinematic plane has thus been probed, where $Q$ is the four-momentum transfer and $x$ the fractional proton energy carried by the struck parton. This continues to be an active field, with the HERA electron-proton collider at DESY producing interesting results at the lowest-$x$ and highest-$Q^2$ values yet reached, and providing new tests of QCD from the jet structure of the hadronic current.

An important theoretical issue is the presence of power-law corrections to the perturbative $Q^2$-evolution, or 'higher-twist' contributions, of the form $1/Q^2$, which enter with *a priori* unknown $x$-dependent coefficients that cannot in general be calculated. In cases where higher-twist contributions are considered, the coefficient is usually extracted by fitting to the data and is correlated with the $\alpha_s$ value thus determined. A corresponding uncertainty on $\alpha_s$ can then be assigned from its dependence on the size of the higher-twist contribution varied within some reasonable range allowed by the data.

## 4.1 Unpolarised Structure Functions

The nucleon structure function can be decomposed into pieces labelled $F_2(x, Q^2)$ and $xF_3(x, Q^2)$, each of which may be written as a sum of *singlet* and *non-singlet* contributions with respect to the quark flavour content, the former having no net flavour. Except at low-$x$, only the $Q^2$ evolution, not the actual shape, of the structure functions is predicted by QCD, and it is different for the singlet and non-singlet components. In the non-singlet case the $Q^2$ evolution enters only via $\alpha_s(Q^2)$, whereas in the singlet case it enters via both $\alpha_s(Q^2)$ and, for $x \leq 0.5$, the gluon distribution function $g(x, Q^2)$. The $Q^2$ dependence of the $F_2$ structure function has been measured in electron, muon, neutrino and antineutrino scattering experiments. The singlet contribution has been determined using deuterium (e, $\mu$) and iron ($\nu$) targets, and the non-singlet contribution using hydrogen (e, $\mu$) targets. The $Q^2$ dependence of the non-singlet $xF_3$ structure function has been measured using neutrino scattering on iron.

A detailed analysis of the $Q^2$-evolution of measurements of $F_2$ was presented several years ago [43], in which a global fit was performed to hydrogen and deuterium data from electron scattering at SLAC and muon scattering at CERN. For the 'non-singlet' region, $x > 0.25$, significant higher-twist contributions were required to fit the data. The effect of these terms was reduced by restricting the fit to high-$Q^2$ bins to yield:

$$\alpha_s(M_Z^2) = 0.113 \pm 0.003(\text{exp.}) \pm 0.004(\text{theor.}) \qquad (11)$$

where the 'experimental error' includes normalisation uncertainties on the data, and uncertainties on the higher-twist and gluon distribution contributions; the theoretical uncertainty is dominated by varying the renormalisation and factorisation scales in the range $0.1 \leq \mu^2/Q^2 \leq 4$.

The NMC Collaboration also determined $\alpha_s$ using their $F_2$ data, but at *low-x* [44]. In this case the $Q^2$ evolution is dominated by the gluon distribution, which must be constrained simultaneously with $\alpha_s$. Higher-twist terms were included as fixed contributions and were not determined from the data. NMC found: $\alpha_s(7 \text{ GeV}^2)$ = 0.264 ± 0.018 (stat.) ±0.070 (syst.) ±0.013 (higher-twist); no uncertainty due to scale variation was assigned. This translates to:

$$\alpha_s(M_Z^2) = 0.117 \pm 0.003(\text{stat.})^{+0.010}_{-0.015}(\text{syst.}) \pm 0.002(\text{higher twist}). \qquad (12)$$

The CCFR Collaboration similarly fitted the $Q^2$-dependence of $xF_3$. They required $Q^2 >15$ GeV$^2$ in order to minimise higher-twist effects, and found [45]: $\alpha_s(M_Z^2) = 0.111 \pm 0.002$ (stat.) $\pm 0.003$ (syst.) $\pm 0.004$ (theor.), where the systematic error is largely due to the uncertainty in the energy calibration of the detector. This result has very recently been updated following a recalibration of the energy scale. A simultaneous fit to $F_2$ and $xF_3$ yields [46]:

$$\alpha_s(M_Z^2) = 0.119 \pm 0.0015(\text{stat.}) \pm 0.0035(\text{syst.}) \pm 0.004(\text{theor.}); \qquad (13)$$

it can be seen that the central value has increased by 0.008 relative to the published result.

CCFR has also studied the $Q^2$-dependence of the integral over $x$ of $xF_3$ that defines the Gross-Llewellyn-Smith sum rule [46]. This quantity is suitable for determination of $\alpha_s$ as it is independent of the gluon distribution, and it has been calculated up to $O(\alpha_s^3)$ in perturbative QCD. The preliminary CCFR measurement is:

$$\alpha_s(M_Z^2) = 0.108^{+0.003}_{-0.005}(\text{stat.}) \pm 0.004(\text{syst.})^{+0.004}_{-0.006}(\text{higher twist}). \qquad (14)$$

An update of this result that reflects the energy recalibration of the detector is eagerly awaited.

Finally, $\alpha_s$ has recently been determined in a theoretical study of the $F_2$ structure function of the proton, using HERA data at low $x$ and large $Q^2$ [47]. In this kinematic region the structure function exhibits 'double scaling', suggesting that a resummation of terms containing powers of $\log Q^2$ and $\log 1/x$ can be compared reliably with the data. Such a resummed calculation has been performed up to next-to-leading order in $\alpha_s$ including all leading and subleading logarithmic terms. Higher-twist corrections are also claimed to be small. A comparison with the 1993 HERA data yields [47]:

$$\alpha_s(M_Z^2) = 0.120 \pm 0.005(\text{exp.}) \pm 0.009(\text{theor.}). \qquad (15)$$

In this case the largest contribution to the 'experimental' error actually derives from uncertainties associated with the parton distribution functions (PDF), and the theoretical uncertainty is dominated by changing the renormalisation and factorisation scales in the range $0.5 < \mu/Q < 2$. Since this technique appears to offer the promise of a reasonably precise determination of $\alpha_s$ that is independent of other methods used in deep-inelastic scattering, one hopes that it will be adopted by the HERA experimental collaborations and applied to their own data so that experimental errors can be properly assessed.

## 4.2 Polarised Structure Functions

In deep-inelastic scattering of polarised leptons on polarised nuclear targets the longitudinal spin structure functions of the proton and neutron can be measured. The Bjorken sum rule is defined by the integral over $x$ of the difference between these structure functions; from the point-of-view of the determination of $\alpha_s$ this quantity is similar to the GLS sum rule and it has also been calculated perturbatively up to $O(\alpha_s^3)$. Data from a variety of experiments have been examined by the E143 Collaboration [48] and the $Q^2$-dependence of the sum rule has been used to determine: $\alpha_s(M_Z^2) = 0.119^{+0.007}_{-0.019}$ (exp.) assuming no higher-twist contribution, and $\alpha_s(M_Z^2) = 0.113^{+0.011}_{-0.035}$ (exp.) when an *ad hoc* parametrisation of such a term was included in the fit; no additional theoretical uncertainties were considered.

A subset of these data has been included in detailed studies of higher-order perturbative QCD contributions to the Bjorken sum rule. Including estimated $O(\alpha_s^4)$ contributions, and neglecting higher-twist effects, the value $\alpha_s(M_Z^2) = 0.122^{+0.005}_{-0.009}$ (exp.) was obtained [49]. Using a QCD sum-rule estimate of the higher-twist coefficient, with a $\pm 50\%$ uncertainty, and assigning an uncertainty of $\pm 10\%$ to the $O(\alpha_s^4)$ coefficient, adds an additional theoretical uncertainty and leads to $\alpha_s(M_Z^2) = 0.118^{+0.007}_{-0.014}$ [49]. A further recent analysis, including a study of the renormalon structure of the Borel transform of the perturbative prediction for the Bjorken sum rule, as well as application of Padé approximants to estimate $O(\alpha_s^4)$ contributions, yielded [50]:

$$\alpha_s(M_Z^2) = 0.116^{+0.003}_{-0.005}(\text{exp.}) \pm 0.003(\text{theor.}) \qquad (16)$$

where the theoretical uncertainty includes contributions from the uncertainty on the Padé estimate, as well as from the higher-twist coefficient uncertainty. This nominally precise result is encouraging, and warrants application of these new theoretical techniques by the experimental collaborations to their own data.

## 4.3 Jet Final States

Hadronic final states in lepton-hadron scattering can be analysed in a similar fashion to those in $e^+e^-$ annihilation (Section 3.3) via infra-red- and collinear-safe measures. The ratio of cross sections $R_{2+1} = \sigma_{2+1}/\sigma_{tot}$, where the '2' refers to events containing two resolved jets in the final state in addition to '+1' proton remnant jet, is particularly suitable since it is directly proportional to $\alpha_s$ and has recently been calculated at next-to-leading order [51] for the JADE jet algorithm [52]. This observable is subject to similar hadronisation uncertainties as those encountered in e$^+$e$^-$ annihilation, and to additional sources of uncertainty relating to the parton distribution functions of the proton as well as contributions to hadronic activity from initial-state parton interactions.

The H1 Collaboration divided their data sample into 5 bins of $Q^2$ in the range $10 < Q^2 < 4000$ GeV$^2$ and evaluated $R_{2+1}$. Using PDFs fitted to HERA data, they compared the QCD prediction with data in the two highest $Q^2$ bins and measured [53]: $\alpha_s(M_Z^2) = 0.123 \pm 0.012$ (stat.) $\pm 0.007$ (syst.) $\pm 0.011$ (theor.). The experimental systematic error is dominated by the uncertainty on the hadronic energy scale of the calorimeter, and the theoretical uncertainty includes contributions from the QCD model-dependence of the acceptance corrections, choice of structure functions, value of $y_c$, and variation of the renormalisation and factorisation scales in the ranges $1/4 \leq \mu^2/Q^2 \leq 4$. A similar analysis by the ZEUS Collaboration, based on 3 $Q^2$ bins in the range $120 < Q^2 < 3600$ GeV$^2$, found [54]: $\alpha_s(M_Z^2) = 0.117 \pm 0.005$ (stat.) $^{+0.004}_{-0.005}$ (syst.) $^{+0.005}_{-0.004}$ (had.) $\pm 0.001$ (PDFs) $^{+0.005}_{-0.006}$ (scale), where variation of the renormalisation and factorisation scales in the range $0.4 \leq \mu^2/Q^2 \leq 2$ was considered. Averaging over these values assuming uncorrelated experimental errors yields:

$$\alpha_s(M_Z^2) = 0.118 \pm 0.006 (\text{exp.}) \pm 0.011 (\text{theor.}). \qquad (17)$$

Interestingly ZEUS and H1 differ in their estimates of theoretical uncertainties from nominally the same sources. For example, from PDFs ZEUS quotes $\pm 0.001$, whereas H1 quotes $\pm 0.005$, and from variation of the renormalisation and factorisation scales ZEUS quotes $^{+0.005}_{-0.006}$ and H1 only $\pm 0.003$, despite the fact that H1 considered a larger range of scale change. Until this situation is resolved the larger H1 estimate of the theoretical uncertainty has been applied to the average value.

Since these are the first such analyses from HERA, and since similar analysis technology has been in use in e$^+$e$^-$ annihilation for over 15 years, there is every reason to expect that the sophistication of the analyses will improve and that the error estimates will stabilise. It will be interesting to see if different HERA jet observables are as sensitive to uncalculated higher-order perturbative contributions as those in e$^+$e$^-$ annihilation, and to what extent they yield a similar degree of scatter in $\alpha_s(M_Z^2)$ (Section 3.3).

# 5 Hadron-hadron Collisions

Hadron-hadron collisions provide a challenging environment for precise measurement of $\alpha_s$. On the experimental side final-state jet measurements are complicated by remnants of the initial-state hadrons, and on the theoretical side calculations at next-to-leading order in perturbation theory include contributions from a large number of Feynman diagrams and have only recently been achieved for jet processes. Finally, the coefficients of powers of $\alpha_s$ in the jet matrix elements depend upon the parton distribution functions of the incoming hadrons; these are constrained at low $Q^2$ scales by deep-inelastic scattering experiments, but uncertainties accrue in their extrapolation to the higher jet transverse energy, $E_T$, scales that are currently accessible at the TeVatron. Any attempt to extract $\alpha_s$ from the jet data is hence affected by these uncertainties, especially in the gluon distribution, and care must be taken to extract simultaneously a common value of $\alpha_s$ used in both the hard-scattering matrix elements and the evolution of the PDFs.

## 5.1 ($W$ + 1-jet)/($W$ + 0-jet) Ratio in p$\overline{\text{p}}$ Collisions

Several experimental collaborations have attempted to determine $\alpha_s$ from the ratio $R$ of the cross sections for production of final states containing a $W$ boson + 1-jet and a $W$ boson + 0-jets; $R$ is proportional to $\alpha_s$ at leading order, and many sources of experimental uncertainty are expected to cancel. The first two studies of this technique were performed at the CERN p$\overline{\text{p}}$ collider before a full next-to-leading order calculation of $R$ had been achieved, and hence relied upon calculations of so-called '$K$-factors' to estimate higher-order QCD corrections to the leading-order result. Data from the UA1 Collaboration were used to measure [55]: $\alpha_s(M_W^2)$ = 0.127 ± 0.026 (stat.) ±0.034 (syst.), where the systematic error is dominated by a contribution of ±0.025 from the $K$-factor uncertainty, with additional large contributions of ±0.013 from PDFs and ±0.010 from fragmentation modelling. The UA2 Collaboration performed a similar analysis, but used next-to-leading order calculations of the total $W$ production cross section as well as the $W$ $p_T$-distribution, to make a more accurate determination of the $K$-factors. They found [56] $\alpha_s(M_W^2)$ = 0.123 ± 0.018 (stat.) ±0.017 (syst.); the systematic error comprises components of ±0.012 from experimental sources and ±0.011 from theoretical uncertainties, the dominant contribution to the latter, ±0.010, arising from fragmentation modelling. A strong sensitivity to the renormalisation scale was found; variation of $\mu$ from $M_W$ to $M_W/2$ caused a change $\Delta\alpha_s = -0.010$, but no corresponding theoretical uncertainty was assigned.

Averaging these results by weighting by the experimental errors assuming they are uncorrelated, and evolving to the scale $M_Z$, yields:

$$\alpha_s(M_Z^2) \;\; = \;\; 0.121 \pm 0.018(\text{exp.}) \pm 0.011(\text{theor.}) \pm 0.010(\text{scale}) \qquad (18)$$

where the UA2 theoretical uncertainty has been taken and the last error has been assigned, based on the UA2 scale-dependence, as an estimate of higher-order uncertainties.

Recently a complete next-to-leading order calculation of $R$ has been performed [57], and has been utilised by the D0 Collaboration in an attempt to measure $\alpha_s$ [58]. Their measurement of $R$ with an accuracy of about $\pm 10\%$ allows in principle a determination of $\alpha_s(M_W^2)$ with an experimental error of $\pm 0.016$. However, they were unable to obtain a value of $\alpha_s$ from a fit of the hard matrix element that is consistent with the value used as input to the parton distribution functions, and concluded that the sensitivity of this technique to $\alpha_s$ is much smaller than expected.

## 5.2 Inclusive Jet Cross Sections

In principle $\alpha_s$ can be determined from inclusive jet cross section measurements, for example the inclusive 1-jet $E_T$ distribution. A recent CDF measurement of the latter observable [59] has aroused much interest due to speculation on the origin of the population of events at high $E_T$. This population is now believed to be explicable by a larger gluonic content of the proton at high fractional momentum than was previously expected [60]; this provides graphic illustration of the necessity to constrain parton distribution functions before other parameters, such as $\alpha_s$, can be determined from the data.

A demonstration of the possibility to determine $\alpha_s$ from the single-jet inclusive $E_T$ distribution has been provided recently [61]. Using a given $\alpha_s$ value as input to a particular parametrisation of PDFs, and fitting the next-to-leading order prediction to the CDF Run 1a data for $E_T > 30$ GeV, it appears possible to measure $\alpha_s(M_Z^2)$ to a precision of $\pm 0.001$ (stat.) $\pm 0.008$ (syst.) $\pm 0.005$ (theor.), where the theoretical uncertainty was defined by variation of the renormalisation scale in the range $0.5 \leq \mu/E_T \leq 2$, and potential non-perturbative effects were not considered. This is an encouraging first step. A programme of further studies is planned, whereby the dependence on the choice of PDFs will be systematiclly studied, and ultimately the PDFs themselves will be extracted simultaneously with $\alpha_s(M_Z^2)$ by fitting to the triply-differential di-jet inclusive distributions [61].

Recently the UA1 Collaboration has released a measurement of $\alpha_s$ based on the cross section for $b\bar{b}$ production. They performed a fit of the next-to-leading order calculation of 'quasi-2-body' $b\bar{b}$ final-states to their data [62] and found: $\alpha_s(20 \text{ GeV})$ = $0.145^{+0.012}_{-0.010}$ (exp.) $\pm 0.003$ ($m_b$) $^{+0.010}_{-0.012}$ (scale) $^{+0.007}_{-0.010}$ (PDF), which can be evolved to:

$$\alpha_s(M_Z^2) = 0.113^{+0.007}_{-0.006}(\text{exp.})^{+0.008}_{-0.009}(\text{theor.}), \tag{19}$$

where the (dominant) renormalisation scale uncertainty corresponds to a variation $1/4 \leq \mu/Q \leq 1$ with $Q = \sqrt{(km_b)^2 + p_t^2}$ and $k$ is an additional fit parameter. This result represents one of the most precise determination of $\alpha_s(M_Z^2)$ in hadronic collisions to date.

## 5.3 Direct Photon Production in pp and p$\overline{\text{p}}$ Collisions

Another reasonably precise determination was provided by the UA6 Collaboration, who measured inclusive direct photon production cross sections in pp and p$\overline{\text{p}}$ collisions using a molecular hydrogen gas-jet target at the CERN p$\overline{\text{p}}$ collider [63]. They were able to isolate the contribution from the process $q\overline{q} \rightarrow \gamma$ g, which has been calculated at next-to-leading order in $\alpha_s$ [64]. Using parton distribution function parametrisations fitted to BCDMS data they determined:

$$\alpha_s(M_Z^2) = 0.112 \pm 0.006(\text{stat.}) \pm 0.005(\text{syst.})^{+0.009}_{-0.001}(\text{theor.}), \tag{20}$$

where the theoretical uncertainty was defined by variation of the renormalisation scale in the range $p_t^2/8 \leq \mu^2 \leq 3p_t^2/4$, the central value being based on the PMS-optimised scale (Section 3.3).

# 6 Heavy Quarkonium Systems

Heavy quarkonium systems can be used to determine $\alpha_s$ either from the measured hadronic decay rate or via the strength of the binding provided by the strong potential; the latter is achieved in practice by comparing the measured energy-level splittings with a lattice QCD calculation.

## 6.1 Heavy Quarkonium Decays

The partial widths $\Gamma_{ggg}$ and $\Gamma_{\gamma gg}$ for the decay of the $^3S$ states of the J/$\Psi$ and $\Upsilon$ into 3 gluons, or a direct photon + 2 gluons, respectively, have been calulated perturbatively at next-to-leading order. By dividing by the partial width $\Gamma_{ll}$ for the leptonic decay the dependence on the wavefunction cancels and $\alpha_s$ can in principle be determined. A non-relativistic analysis [65] revealed large differences between the $\Lambda_{\overline{MS}}$ values determined in this fashion from the J/$\Psi$ and $\Upsilon$ systems, as well as a large renormalisation scale dependence. An improved analysis incorporated *ad hoc* relativistic corrections as well as next-to-next-to-leading order (NNLO) terms that were determined by fitting simultaneously to the data on $\Gamma_{ggg}/\Gamma_{ll}$ for the J/$\Psi$ and $\Upsilon$ 1S - 3S resonances. The relativistic corrections turn out to be large for the J/$\Psi$. A preliminary result: $\alpha_s(M_Z^2) = 0.113 \pm 0.001$ (exp.) $^{+0.007}_{-0.005}$ (theor.) was obtained [65], where the theoretical uncertainty was defined from the variation of the fitted NNLO term for a scale variation in the range $0.5 \leq \mu/Q \leq 2$; a final version of this analysis has yet to be published.

CLEO similarly used their measured ratio $\Gamma_{\gamma gg}/\Gamma_{ggg}$ for the $\Upsilon$(1S) and have presented a preliminary result [12] $\alpha_s(M_Z^2) = 0.110 \pm 0.001$ (exp.) $\pm 0.004$ (syst.) $\pm 0.005$ (theor.), where the last uncertainty was defined by $0.1 \leq \mu^2/Q^2 \leq 1$. Combining with the analysis of [65], assuming uncorrelated experimental errors, yields a preliminary average:

$$\alpha_s(M_Z^2) = 0.113 \pm 0.001(\text{exp.})^{+0.007}_{-0.005}(\text{theor.}). \tag{21}$$

## 6.2 Lattice Gauge Theory

Lattice gauge theory currently provides a successful tool for performing non-perturbative QCD calculations, although it is presently limited in applicability to static properties of hadrons. Experimental data on hadron properties, such as meson masses or energy-level splittings in heavy mesons, can be input to lattice simulations and $\alpha_s$ extracted by calculation; for a recent review see [66]. Currently the precision of such determinations of $\alpha_s(M_Z^2)$ is not limited by experiment, but by the various theoretical uncertainties relating to lattice discretisation, treatment of 'sea quarks', and matching between the different renormalisation schemes used in lattice and perturbative calculations.

Considerable interest was generated several years ago with the determination by the FNAL/SCRI group of a nominally very precise $\alpha_s(M_Z^2)$ measurement using the 1P - 1S splitting in the charmonium system: $\alpha_s(M_Z^2) = 0.105 \pm 0.004$ [67]. The dominant error was derived from extrapolating from the 'quenched approximation', corresponding to zero quark flavours in vacuum polarisation contributions to the strong coupling, to four-flavour contributions above the charmonium energy scale. A somewhat larger, and even more precise, result, based on input data from the $\Upsilon$ spectrum, was presented more recently by the NRQCD group [68]: $\alpha_s(M_Z^2) = 0.115 \pm 0.002$. In this case both zero and two flavours of dynamical quarks were considered, and the result extrapolated to three flavours, before evolving through the $c\bar{c}$ and $b\bar{b}$ mass thresholds up to $M_Z$. The dominant error was derived from the uncertainty in the two-loop matching between the $\alpha_V$ and $\overline{MS}$ schemes. An independent determination based on the $\rho$ mass and the charmonium 1S-1P splitting, and using two flavours of dynamical quarks, yielded [69]: $\alpha_s(M_Z^2) = 0.108 \pm 0.008$. A review of these measurements quoted [66]: $\alpha_s(M_Z^2) = 0.112 \pm 0.007$, the conservative error estimate reflecting a dispersion among the values from the different groups that is larger than the quoted errors in some cases.

Further progress has been made recently; the two-loop matching coefficient has been calculated for the quenched approximation, and errors relating to the dynamical quark mass value have been considered. New preliminary $\alpha_s(M_Z^2)$ results have been presented [70] by the FNAL/SCRI group: $0.116 \pm 0.003$, and by the NRQCD group: $0.118 \pm 0.003$, where in both cases the two-loop matching uncertainty is dominant. This recent convergence of the lattice results is encouraging, although the shifts in the central values of $\alpha_s(M_Z^2)$, as well as changes in the error estimates, as the techniques have improved is perhaps an indication that the systematics are not yet fully under control. An average over the two recent preliminary results yields:

$$\alpha_s(M_Z^2) = 0.117 \pm 0.003 (\text{theor.}) \qquad (22)$$

where the error is largely from common theoretical uncertainties.

## 7 Summary and Outlook

The average $\alpha_s(M_Z^2)$ value from each method, derived as described in the text, is shown in Table 1, together with the total experimental error and theoretical uncertainty. For the benefit of the aesthetically-minded the same data are summarised in Figure 1. Perhaps the most difficult aspect of this review is to combine these 17 results in a meaningful way. It is worth restating that, in all cases, except perhaps the $Z^0$ lineshape measurement, the theoretical uncertainty is a non-negligible contribution to the total error, and in many cases it is larger than the experimental error. By definition such theoretical uncertainties can only be estimated in an *ad hoc* fashion, and the degree of conservatism of the assigned uncertainty can often be contested. I have attempted to summarise the contributions to the theoretical uncertainty as quoted by the authors of the respective measurements and to assess the 'degree of reasonableness' of their estimate. In those cases where subsequent developments suggest a different estimate than that originally assigned, I have changed the theoretical uncertainty accordingly; in all cases the original results are included in the text and can be recovered and treated differently by those who hold a different view.

It is certainly not correct to assume that all 17 measurements of $\alpha_s(M_Z^2)$ summarised in Table 1 are completely independent of one another. Correlations between measurements and between the systematic errors, especially on the theoretical side, certainly exist. However, no attempt has previously been made in a review of this kind to evaluate the correlations between the different measurements and to treat the errors accordingly in forming an average; this tradition will not be broken here. Taking an average over all 17 measurements *assuming* they are independent, by weighting each by its *total* error, yields $\alpha_s(M_Z^2) = 0.118$ with a $\chi^2$ of 6.4; the low $\chi^2$ value reflects the fact that most of the measurements are systematics-limited. Taking an unweighted average, which in some sense corresponds to the assumption that all 17 measurements are completely correlated, yields the same result. There can hence, hopefully, be little dispute that the average value $\alpha_s(M_Z^2) = 0.118$ accurately characterises the 17 measurements.

The first procedure yields an error on the average of $\pm 0.001$; because of correlations this is almost certainly an underestimate of the 'true error'. In the spirit of the second procedure one could take the error on the most precise measurement, namely $\pm 0.003$; this has been adopted by a previous reviewer [3]. Instead I choose to quote the r.m.s. deviation of the 17 measurements, w.r.t. their average value, to characterise the dispersion:

$$\alpha_s(M_Z^2) = 0.118 \pm 0.005, \tag{23}$$

which is presumably a conservative estimate of the error. In an independent review [34] an attempt was made to consider positive correlations between measurements by rescaling errors according to the respective $\chi^2$ value. This yielded the same average $\alpha_s(M_Z^2)$ value as presented here, and an error of $\pm 0.003$.

It is self-evident from Fig. 1 that, within the errors, there is no evidence of any discrepancy between measurements made at 'low'- and 'high'-$Q^2$. It is notable, however, that the precision of the individual measurements varies between about 3 and 20%, so that any anomalous effects up to the 5% level would not necessarily be discerned.

If further progress is to be made in testing QCD, future measurements of $\alpha_s(M_Z^2)$ should aim for substantially improved precision. The prospects for achieving 1%-level measurements are discussed elsewhere [71]. Lattice QCD determinations may reach this precision within the next few years. Deep-inelastic scattering and $e^+e^-$ annihilation will probably require higher-energy facilities, as well as significant theoretical effort in the areas of higher-twist and $O(\alpha_s^3)$ perturbative contributions, respectively. A precise $\alpha_s(M_Z^2)$ measurement has yet to emerge from the TeVatron, but feasibility studies are in progress and appear promising.

## Acknowledgements

Many people have helped to reduce my ignorance during the preparation of this article. I am particularly grateful to S. Brodsky, T. Browder, A. El Khadra, W. Gary, W. Giele, N. Glover, D. Harris, T. Rizzo and B. Schumm for helpful conversations.

# References

[1] H. Fritzsch, M. Gell-Mann, H. Leutwyler, Phys. Lett. **B47** (1973) 365; D.J. Gross, F. Wilczek, Phys. Rev. Lett. **30** (1973) 1343; H.D. Politzer, Phys. Rev. Lett. **30** (1973) 1346.

[2] W.A. Bardeen, A.J. Buras, D.W. Duke, T. Muta, Phys. Rev. **18** (1978) 3998; D.W. Duke, Rev. Mod. Phys. **52** (1980) 199.

[3] R.M. Barnett *et al.*, Phys. Rev. **D54** (1996) 77.

[4] R.M. Barnett *et al.*, Phys. Rev. **D54** (1996) 19.

[5] G.L. Kane *et al.*, Phys. Lett. **B354** (1995) 350.

[6] B.R. Webber, Proc. XXVII International Conference on High Energy Physics, July 20-27 1994, Glasgow, Scotland, IoP Publishing, Eds. P.J. Bussey, I.G. Knowles, p. 213; S. Bethke, Nucl. Phys. B (Proc. Suppl.) **39B**, C (1995) 198.

[7] See *eg.* M. Beneke, SLAC-PUB-7277 (1996); to appear in Proc. 28th International Conference on High Energy Physics, 25-31 August 1996, Warsaw, Poland;

[8] P.N. Burrows *et al.*, Phys. Lett. **B382** (1996) 157.

[9] S.G. Gorishny, A. Kataev, S.A. Larin, Phys. Lett. **B259** (1991) 144; L.R. Surguladze, M.A. Samuel, Phys. Rev. Lett. **66** (1991) 560.

[10] K.H. Chetyrkin, J.H. Kühn, A. Kwiatkowski, Berkeley preprint LBL 36678-Rev (1996); subm. to Phys. Rep.

[11] W. de Boer, R. Ehret, S. Schael, Proc. of the workshop on QCD at LEP, Aachen, April 11 1994; PITHA 94/33 (1994).

[12] T. Browder, 'Experimental results on QCD from CLEO II', to appear in Proc. 28th International Conference on High Energy Physics, 25-31 August 1996, Warsaw, Poland.

[13] A. Blondel, 'Experimental status of electroweak interactions', to appear in Proc. 28th International Conference on High Energy Physics, 25-31 August 1996, Warsaw, Poland.

[14] P.B. Renton, Proc. 17th International Symposium on Lepton-Photon Interactions, 10-15 August 1995, Beijing, China, p. 35.

[15] D. Schaile, Proc. XXVII International Conference on High Energy Physics, July 20-27 1994, Glasgow, Scotland, IoP Publishing, Eds. P.J. Bussey, I.G. Knowles, p. 27.

[16] See *eg.* M. Shifman, Mod. Phys. Lett. **A10** No. 7 (1995) 605.

[17] J. Chyla, A.L. Kataev, Reports of the Working Group on Precision Calculations for the Z Resonance, eds. D. Bardin, W. Hollik, G. Passarino, CERN 95-03 (1995) 313.

[18] F. Le Diberder, A. Pich, Phys. Lett. **B289** (1992) 165.

[19] E. Braaten, S. Narison, A. Pich, Nucl. Phys. **B373** (1992) 581; F. Le Diberder, A. Pich, Phys. Lett. **B286** (1992) 147.

[20] M. Neubert, Nucl. Phys. **B463** (1996) 511.

[21] ALEPH Collab., D. Buskulic *et al.*, Phys. Lett. **B307** (1993) 209.

[22] L. Duflot (ALEPH Collab.), Nucl. Phys. B (Proc. Supp.) **39B** (1995) 322.

[23] OPAL Collab., R. Akers *et al.*, Z. Phys. **C66** (1995) 543.

[24] CLEO Collab., T. Coan *et al.*, Phys. Lett. **B356** (1995) 580.

[25] G. Altarelli, P. Nason, G.Ridolfi, Z. Phys. **C68** (1995) 257.

[26] SLD Collab., K. Abe *et al.*, Phys. Rev. Lett. **71** (1993) 2528.

[27] SLD Collab., K. Abe *et al.*, Phys. Rev. **D51** (1995) 962.

[28] R.K. Ellis, D.A. Ross, A.E. Terrano, Phys. Rev. Lett. **45** (1980) 1226; Nucl. Phys. **B178** (1981) 421; G. Kramer, B. Lampe, Z. Phys. **C39** (1988) 101; Fortschr. Phys. **37** (1989) 161; Z. Kunszt *et al.*, CERN 89–08 Vol I, (1989) p. 373.

[29] T. Sjöstrand, CERN–TH–7112/93 (1993).

[30] G. Marchesini *et al.*, Comp. Phys. Comm. **67** (1992) 465.

[31] S. Catani, G. Turnock, B.R. Webber, and L. Trentadue, Phys. Lett. **B263** (1991) 491.

[32] ALEPH: D. Decamp *et al*, Phys. Lett. **B284** (1992) 163; L3: O. Adriani *et al*, Phys. Lett. **B284** (1992) 471; OPAL: P.D. Acton *et al*, Z. Phys. **53** (1993) 1; DELPHI: P. Abreu *et al*, Z. Phys. **53** (1993) 21.

[33] D. Duchesneau, 'QCD Results from LEP above and below the $Z^0$ Peak', to appear in Proc. 28th International Conference on High Energy Physics, Warsaw, Poland, 25-31 July 1996.

[34] M. Schmelling, 'Status of the Strong Coupling Constant', to appear in Proc. 28th International Conference on High Energy Physics, Warsaw, Poland, 25-31 July 1996.

[35] P.M. Stevenson, Phys. Rev. **D23** (1981) 2916; G. Grunberg, Phys. Rev. **D29** (1984) 2315; S.J. Brodsky, G.P. Lepage, P.B. Mackenzie, Phys. Rev. **D28** (1983) 228.

[36] P.N. Burrows *et al.*, SLAC-PUB-7222 (1996); subm. to Phys. Lett. B.

[37] L3 Collab., S. Banerjee *et al.*, paper contributed to 28th International Conference on High Energy Physics, Warsaw, Poland, 25-31 July 1996, pa04-021.

[38] DELPHI Collab., P. Abreu *et al.*, Phys. Lett. **B311** (1993) 408.

[39] ALEPH Collab., D. Buskulic *et al.*, Phys. Lett. **B357** (1995) 487.

[40] G. Gurci, W. Furmanski, R. Petronzio, Nucl. Phys. **B175** (1980) 27.

[41] P. Nason, B.R. Webber, Nucl. Phys. **B421** (1994) 473.

[42] DELPHI Collab., W. de Boer *et al.*, paper contributed to 28th International Conference on High Energy Physics, Warsaw, Poland, 25-31 July 1996, pa01-022.

[43] M. Virchaux, A. Milsztajn, Phys. Lett. **B274** (1992) 221.

[44] NMC Collab., M. Arneodo *et al.*, Phys. Lett. **B309** (1993) 222.

[45] CCFR Collab., P.Z. Quintas *et al.*, Phys. Rev. Lett. **71** (1993) 1307.

[46] D. Harris, 'New results from the CCFR collaboration', to appear in Proc. 28th International Conference on High Energy Physics, Warsaw, Poland, 25-31 July 1996.

[47] R.D. Ball, S. Forte, Phys. Lett. **B358** (1995) 365.

[48] E143 Collab., K. Abe *et al.*, Phys. Lett. **B364** (1995) 61.

[49] J. Ellis, M. Karliner, Phys. Lett. **B341** (1995) 397.

[50] J. Ellis *et al.*, Phys. Lett. **B366** (1996) 268.

[51] D. Graudenz, Phys. Rev. **49** (1994) 3291.

[52] JADE Collab., W. Bartel *et al.*, Z. Phys. **C33** (1986) 23.

[53] H1 Collab., T. Ahmed *et al.*, Phys. Lett. **B346** (1995) 415.

[54] ZEUS Collab., M. Derrick *et al.*, Phys. Lett. **B363** (1995) 201.

[55] M. Lindgren *et al.*, Phys. Rev. **D45** (1992) 3038.

[56] UA2 Collab., J. Alitti *et al.*, Phys. Lett. **B263** (1991) 563.

[57] W.T. Giele, E.W.N. Glover, D.A. Kosower, Nucl. Phys. **B403** (1993) 633.

[58] D0 Collab., S. Abachi *et al.*, Phys. Rev. Lett. **75** (1995) 3226.

[59] CDF Collab., F. Abe *et al.*, Phys. Rev. Lett. **77** (1996) 438.

[60] J. Huston *et al.*, Phys. Rev. Lett. **77** (1996) 444.

[61] W.T. Giele, E.W.N. Glover, J. Yu, Phys. Rev. **D53** (1996) 120.

[62] UA1 Collab., C. Albajar *et al.*, Phys. Lett. **B369** (1996) 46.

[63] UA6 Collab., G. Sozzi *et al.*, Phys. Lett. **B317** (1993) 243.

[64] P. Aurenche *et al.*, Nucl. Phys. **B297** (1988) 661.

[65] M. Kobel, 'Perturbative QCD and Hadronic Interactions', Proc. XXVII Rencontres de Moriond, 22-28 March 1992, Les Arcs, Savoie, France, Editions Frontieres (1992), ed. J.Tran Thanh Van, p.145.

[66] C. Michael, Nucl. Phys. B (Proc. Suppl.) **42** (1995) 147.

[67] A.X. El Khadra *et al.*, Phys. Rev. Lett. **69** (1992) 729.

[68] C.T.H. Davies *et al.*, Phys. Lett. **B345** (1995) 42.

[69] S. Aoki *et al.*, Phys. Rev. Lett. **74** (1995) 22.

[70] J. Shigemitsu, 'Quarkonium physics and $\alpha_{strong}$ from quarkonia', Ohio State preprint OHSTPY-HEP-T 96-015 (1996); to appear in Proc. 14th International Symposium on Lattice Field Theory, 4-8 June 1996, St. Louis, Mo, USA.

[71] P. Burrows *et al.*, 'Prospects for the precision measurement of $\alpha_s$ ', to appear in Proc. Workshop on future directions in high energy physics, June 25 - July 12 1996, Snowmass, Co, USA.

| Method | $Q$ (GeV) | $\alpha_s(M_Z^2)$ | exp. | theor. |
|---|---|---|---|---|
| e$^+$e$^-$: $\tau$ decays | 1.8 | 0.122 | 0.001 | 0.006 |
| DIS: Bjorken SR | 1 - 4 | 0.116 | $^{+0.003}_{-0.005}$ | 0.003 |
| DIS: GLS SR | 1 - 5 | 0.108 | $^{+0.005}_{-0.006}$ | $^{+0.004}_{-0.006}$ |
| DIS: $F_2$ (NMC) | 1 - 7 | 0.117 | $^{+0.010}_{-0.015}$ | 0.002 |
| DIS: $F_2$ (HERA) | 1 - 10 | 0.120 | 0.005 | 0.009 |
| DIS: $F_2$ (SLAC, BCDMS) | 2 - 16 | 0.113 | 0.003 | 0.004 |
| pp, p$\overline{\text{p}}$: direct $\gamma$ | 4 - 6 | 0.113 | 0.008 | $^{+0.009}_{-0.001}$ |
| LGT: J/$\Psi$, $\Upsilon$ | 3 - 11 | 0.117 | — | 0.003 |
| J/$\Psi$, $\Upsilon$ decays | 3 - 11 | 0.113 | 0.001 | $^{+0.007}_{-0.005}$ |
| DIS: $F_2$, $xF_3$ (CCFR) | 3 - 23 | 0.119 | 0.004 | 0.004 |
| DIS: jets | 10 - 63 | 0.118 | 0.006 | 0.011 |
| p$\overline{\text{p}}$: b$\overline{\text{b}}$ prod. | 20 | 0.113 | $^{+0.007}_{-0.006}$ | $^{+0.008}_{-0.009}$ |
| e$^+$e$^-$: $R$ | 5 - 65 | 0.128 | $^{+0.012}_{-0.013}$ | 0.002 |
| p$\overline{\text{p}}$: W+1-jet | 80 | 0.121 | 0.018 | 0.015 |
| e$^+$e$^-$: event shapes | 10 - 133 | 0.121 | 0.002 | 0.005 |
| e$^+$e$^-$: fragmentn. fns. | 14 - 91 | 0.124 | 0.005 | 0.010 |
| e$^+$e$^-$: $Z^0$ lineshape | 91 | 0.1202 | 0.0033 | $< 0.001$ |

Table 1: Average $\alpha_s(M_Z^2)$ values and errors from the 17 methods described in the text. The approximate range in $Q$ is shown for each method.

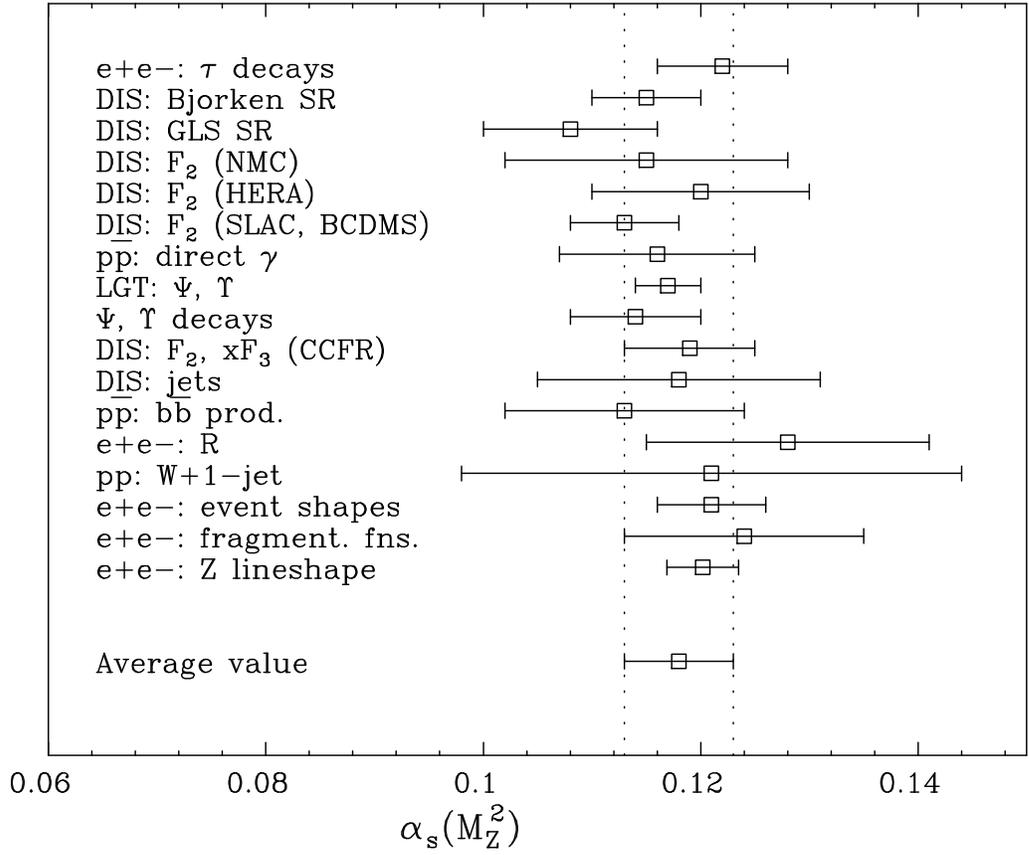

Figure 1: Average $\alpha_s(M_Z^2)$ values and errors from the 17 methods described in the text. The results are ordered vertically in $Q$.

28